\documentclass[12pt,a4paper]{article}

\usepackage[utf8]{inputenc}
\usepackage[T1]{fontenc}
\usepackage{amsmath,amssymb}
\usepackage{geometry}
\usepackage{hyperref}
\title{\bf Wilson loops as probes of phase transitions and conductivity phenomena}
\author{
Tetiana Obikhod$^{1}$, Ievgenii Petrenko\\[2mm]
$^{1}$Institute for Nuclear Research NAS of Ukraine, Kyiv, Ukraine
}
\date{}

\begin{document}
\maketitle

\begin{abstract}
Wilson loops are among the most fundamental gauge-invariant observables in quantum field theory, encoding the global structure of gauge fields through their holonomy along closed contours. Originally introduced as order parameters for confinement in non-Abelian gauge theories, they have recently acquired a central role in condensed matter physics, where they characterize topological phases and quantized transport phenomena. In this work we present a unified theoretical picture in which Wilson loops connect nonperturbative gauge dynamics, Berry-phase topology in band theory, and the quantum Hall response of interacting electron systems. We demonstrate explicitly how Wilson loops encode Chern numbers, fractional charge, and anyonic braiding statistics within Chern--Simons effective field theory. Both quantized Hall conductivity and quasiparticle statistics are shown to originate from the same topological invariant --- the linking number of Wilson loops --- establishing a direct correspondence between microscopic topological structure and macroscopic transport.
\end{abstract}

\section{Introduction}

Nonlocal gauge-invariant observables occupy a central place in modern theoretical physics. Among them, Wilson loops stand out as particularly powerful probes of the global structure of gauge fields. Introduced originally by Wilson in the context of non-Abelian gauge theories as order parameters for confinement~[1], they encode the holonomy of gauge connections along closed contours and thereby probe information inaccessible to local operators. Their area-law or perimeter-law behavior sharply distinguishes confining and deconfining phases and forms the conceptual foundation of lattice gauge theory and nonperturbative quantum chromodynamics~[2].

Over the last decades, Wilson loops have found a new and unexpected arena in condensed matter physics. In topological phases of matter --- including quantum Hall states, topological insulators, and interacting topologically ordered systems --- the organizing principle is no longer spontaneous symmetry breaking but the global geometry and topology of quantum states in Hilbert space~[3,4,5]. These phases exhibit quantized response coefficients, robust boundary modes, and exotic quasiparticle statistics, all of which are intrinsically nonlocal and insensitive to microscopic details. Wilson loops provide a natural and unifying language for describing these phenomena through Berry phases, Chern numbers, and braiding holonomies~[6,7].

The fractional quantum Hall effect (FQHE) represents perhaps the clearest realization of intrinsic topological order. It displays fractional electric charge, anyonic exchange statistics, and topological ground-state degeneracy on manifolds of nontrivial topology~[3,5]. These features cannot be understood within the conventional Landau paradigm, but follow naturally from effective Chern--Simons gauge theories~[6,7], in which Wilson loops represent quasiparticle worldlines and encode their mutual statistics via topological linking numbers.

The aim of this paper is to develop a coherent framework in which Wilson loops play a unifying role across these diverse physical settings. We show how they diagnose phase structure in gauge theories, encode Berry-phase topology in band theory, and govern transport and statistics in quantum Hall systems. In particular, we demonstrate that in the fractional quantum Hall regime the same Wilson loops simultaneously determine both the quantized Hall conductivity and the anyonic braiding phases, revealing transport and statistics as dual manifestations of a single underlying topological invariant~[3,6,7].

\section{Wilson loops as probes of gauge structure and topology}

We begin by recalling the definition and physical meaning of Wilson loops in gauge theory. Let
\begin{equation}
A_\mu(x)=A_\mu^a(x)T^a
\end{equation}
be a gauge connection valued in the Lie algebra of a gauge group $G$. For a closed contour $C$, the Wilson loop operator is defined as
\begin{equation}
W(C)=\frac{1}{\dim R}\,\mathrm{Tr}_R\,\mathcal{P}\exp\!\left(i\oint_C A_\mu(x)\,dx^\mu\right),
\end{equation}
where $\mathcal{P}$ denotes path ordering and $R$ is a representation of $G$~[1,2]. Under gauge transformations,
\begin{equation}
A_\mu \rightarrow gA_\mu g^{-1}-i(\partial_\mu g)g^{-1},
\end{equation}
the Wilson loop transforms by conjugation, and its trace is therefore gauge invariant. Physically, $W(C)$ measures the holonomy acquired by a test charge transported adiabatically along the loop $C$.

In non-Abelian gauge theories, the expectation value of large Wilson loops provides a diagnostic of confinement~[1,2]. One finds generically
\begin{equation}
\langle W(C)\rangle \sim 
\begin{cases}
\exp[-\sigma\,\mathrm{Area}(C)], & \text{confining phase},\\[4pt]
\exp[-\mu\,\mathrm{Perimeter}(C)], & \text{deconfined phase},
\end{cases}
\end{equation}
where $\sigma$ is the string tension. An area law implies a linear potential between static charges,
\begin{equation}
V(R)\sim \sigma R,
\end{equation}
signaling confinement~[2]. Thus, Wilson loops encode nonperturbative phase structure through global holonomy rather than local order parameters.

In topological gauge theories, such as Chern--Simons theory~[6,7], Wilson loops take on an even more striking role. Their expectation values no longer depend on geometric properties such as the area enclosed by the loop, but only on topological invariants of the loop configuration --- linking numbers and knot invariants~[7]. This purely topological dependence anticipates their relevance to condensed matter systems with intrinsic topological order~[3,4].

\section{Wilson loops, Berry phases, and quantized transport}

A closely related notion of holonomy appears in the theory of Berry phases in quantum mechanics and solid-state physics~[8]. In crystalline solids, Bloch bands define vector bundles over the Brillouin zone. For a nondegenerate band with Bloch eigenstates $|u_n(\mathbf{k})\rangle$, the Berry connection is
\begin{equation}
\mathbf{A}_n(\mathbf{k})= i\langle u_n(\mathbf{k})|\nabla_{\mathbf{k}}|u_n(\mathbf{k})\rangle,
\end{equation}
and the Berry phase accumulated along a closed loop $\mathcal{C}$ in momentum space is
\begin{equation}
\gamma_n(\mathcal{C})=\oint_{\mathcal{C}} \mathbf{A}_n(\mathbf{k})\cdot d\mathbf{k}.
\end{equation}
This is precisely the Wilson loop of the Berry connection~[8],
\begin{equation}
W_n(\mathcal{C})=\exp(i\gamma_n(\mathcal{C})).
\end{equation}

The associated Berry curvature,
\begin{equation}
\mathbf{F}_n(\mathbf{k})=\nabla_{\mathbf{k}}\times \mathbf{A}_n(\mathbf{k}),
\end{equation}
gives rise to the first Chern number of a filled band~[8],
\begin{equation}
C_n=\frac{1}{2\pi}\int_{\mathrm{BZ}} d^2k\, F_{n,z}(\mathbf{k}),
\end{equation}
which is quantized to integer values. By Stokes' theorem,
\begin{equation}
\gamma_n(\partial S)=\int_S \mathbf{F}_n\cdot d\mathbf{S},
\end{equation}
so the quantization of the Chern number corresponds directly to the winding of Wilson loop phases around noncontractible cycles of the Brillouin torus.

This topological structure manifests itself experimentally in the integer quantum Hall effect~[5]. For completely filled bands, the Hall conductivity is given by
\begin{equation}
\sigma_{xy}=\frac{e^2}{h}\sum_n C_n,
\end{equation}
demonstrating that quantized transport arises directly from the topology of Berry-phase Wilson loops~[5,8].

\section{Chern--Simons theory, Wilson loops, and the fractional quantum Hall effect}

The fractional quantum Hall effect (FQHE) reveals a deeper layer of topological physics, beyond the single-particle Berry-phase picture~[3,5]. At low energies, Laughlin states at filling fraction
\begin{equation}
\nu=\frac{1}{m}, \qquad m\in 2\mathbb{Z}+1,
\end{equation}
are described by an effective Abelian Chern--Simons theory~[6,7],
\begin{equation}
S_{\mathrm{CS}}=
\frac{m}{4\pi}\int d^3x\,\epsilon^{\mu\nu\rho}a_\mu\partial_\nu a_\rho
+\frac{e}{2\pi}\int d^3x\,\epsilon^{\mu\nu\rho}A_\mu\partial_\nu a_\rho,
\end{equation}
where $a_\mu$ is an emergent gauge field encoding collective degrees of freedom and $A_\mu$ is the external electromagnetic field.

Integrating out the emergent field $a_\mu$ yields the effective electromagnetic response~[6],
\begin{equation}
S_{\mathrm{eff}}[A]=
\frac{\nu e^2}{4\pi}\int d^3x\,\epsilon^{\mu\nu\rho}A_\mu\partial_\nu A_\rho,
\end{equation}
from which the induced current follows as
\begin{equation}
j^\mu=\frac{\delta S_{\mathrm{eff}}}{\delta A_\mu}
=\frac{\nu e^2}{2\pi}\epsilon^{\mu\nu\rho}\partial_\nu A_\rho,
\end{equation}
implying the quantized Hall conductivity
\begin{equation}
\sigma_{xy}=\nu\frac{e^2}{h}.
\end{equation}

In this framework, quasiparticles correspond to Wilson lines of the emergent gauge field~[3,7],
\begin{equation}
W(C)=\exp\!\left(iq\oint_C a_\mu dx^\mu\right),
\end{equation}
where for Laughlin states the minimal quasiparticle charge is
\begin{equation}
q=\frac{e}{m}.
\end{equation}
The true topological content of the theory emerges when one considers the braiding of such quasiparticles. For two worldlines $C_1$ and $C_2$, the Chern--Simons path integral yields~[7]
\begin{equation}
\langle W(C_1)W(C_2)\rangle
\propto \exp\!\left(i\frac{2\pi}{m}\,\mathrm{Lk}(C_1,C_2)\right),
\end{equation}
where $\mathrm{Lk}(C_1,C_2)$ is the Gauss linking number. A single exchange corresponds to $\mathrm{Lk}=1$ and produces the statistical phase
\begin{equation}
\theta=\frac{\pi}{m},
\end{equation}
demonstrating that Laughlin quasiparticles obey fractional, anyonic statistics~[3,7].

Remarkably, the same topological parameter $m$ governs both transport and statistics~[3,6,7]:
\begin{equation}
\sigma_{xy}=\frac{e^2}{mh}, \qquad \theta=\frac{\pi}{m}.
\end{equation}
Both quantities arise from Wilson loops of the same emergent gauge field, showing that fractional charge, fractional statistics, and quantized conductivity are unified manifestations of a single underlying topological structure.

\section{Wilson loops and intrinsic topological order}

Beyond transport and quasiparticle statistics, Wilson loops also encode the global structure of the ground-state manifold. On a torus, Chern--Simons theory predicts a ground-state degeneracy~[6,7]
\begin{equation}
\mathcal{D}=m,
\end{equation}
which can be understood from the algebra of Wilson loops wrapping the two noncontractible cycles:
\begin{equation}
W_x W_y = e^{2\pi i/m} W_y W_x.
\end{equation}
This noncommutative algebra reflects intrinsic topological order and cannot be attributed to any local order parameter or broken symmetry~[3,4]. Instead, it expresses the fundamentally global nature of the quantum state, once again encoded in Wilson loop holonomies.

\section{Discussion and conclusions}

In this work we have argued that Wilson loops provide a unifying conceptual and mathematical framework connecting confinement in gauge theories, Berry-phase topology in band theory, and topological order in quantum Hall systems. In all these contexts, Wilson loops encode global holonomy --- quantities insensitive to local perturbations yet capable of sharply distinguishing distinct quantum phases.

In topological band theory, Wilson loops of the Berry connection encode Chern numbers and polarization invariants, directly determining quantized transport coefficients. In interacting quantum Hall systems, Wilson loops of emergent gauge fields represent quasiparticle worldlines and encode anyonic braiding statistics through linking numbers. The fact that the same topological invariant controls both Hall conductivity and exchange phases demonstrates that transport and statistics are dual manifestations of a deeper underlying topological structure.

More broadly, these results reinforce the principle that topology stands on equal footing with symmetry as an organizing concept in quantum matter. Wilson loops exemplify this principle by providing observables that diagnose phases without reference to local order parameters. Their extension to non-Abelian Chern--Simons theories opens the door to richer braiding structures and potential applications to topological quantum computation, while recent experimental progress in engineered quantum systems offers promising platforms for probing Wilson-loop physics directly.

Future directions include the study of higher-form symmetries, fracton phases, and higher-order topological insulators using generalized Wilson operators, as well as the systematic classification of interacting topological phases through loop and surface holonomies. In all such contexts, Wilson loops are likely to remain indispensable tools for uncovering the deep geometric and topological structures underlying quantum matter.

\end{document}